\newcommand{\Hcal}{\ensuremath{\mathcal{H}}}
\newcommand{\R}{\mathbb{ R}}
\newcommand{\rd}{\mathrm{d}}

\documentclass[reprint,nofootinbib,superscriptaddress]{revtex4-2}

\usepackage{graphicx} 
\usepackage{dcolumn} 
\usepackage{bm} 
\usepackage{multirow}
\usepackage{booktabs}
\usepackage{amsmath}
\usepackage{hyperref}        
\usepackage[capitalise]{cleveref}

\usepackage{amsthm,amssymb, physics}
\usepackage{xcolor}
\usepackage{enumitem}
\usepackage{soul}
\usepackage{url}
\usepackage[normalem]{ulem}

\usepackage{mathtools}

\newtheorem{Theorem}{Theorem}
\newtheorem{Proposition}{Proposition}
\newtheorem{Lemma}{Lemma}

\newtheorem{definition}{Definition}

\newcommand{\qmu}[0]{\ensuremath{q_{\textrm{MU}}}}

\newcommand{\qbm}[0]{\ensuremath{q_{\textrm{BBM}}}}

\newcommand{\cket}[1]{|#1\rangle}

\begin{document}

\title{Tight Quasiprobabilistic Entropic Uncertainty Relations for Discrete Quantum Systems}

\author{Jonathan J. Thio}
\affiliation{Cavendish Lab., Department of Physics, Univ. of Cambridge, Cambridge CB3 0HE, UK}

\author{Stephan De Bièvre}
\affiliation{Univ. Lille, CNRS, Inria, UMR 8524, Laboratoire Paul Painlevé, F-59000 Lille, France}

\author{Crispin H. W. Barnes}
\affiliation{Cavendish Lab., Department of Physics, Univ. of Cambridge, Cambridge CB3 0HE, UK}

\author{David R. M. Arvidsson-Shukur}
\affiliation{Hitachi Cambridge Lab., J.J. Thomson Avenue, Cambridge CB3 0HE, UK}

\date{\today} 

\begin{abstract}
It was conjectured by Wehrl and proven by Lieb that the Shannon differential entropy of the Husimi function of a continuous-variable quantum state is lower bounded by the value it takes on coherent states. This corroborates the intuition, familiar from Heisenberg's uncertainty relations, that coherent states are minimal uncertainty states. The analogous statement for the Wigner function would be that the Wigner differential entropy of states with a positive Wigner function is bounded below by the entropy of pure Gaussian states. This was conjectured a decade ago, but recently disproven.  
Here, we prove the analogous statement for Kirkwood-Dirac distributions with mutually unbiased bases and for the Gross-Wigner distributions. For these discrete systems, the entropy-minimizing role is played by the pure states with positive quasiprobability distributions. Further, we prove that Kirkwood-Dirac distributions with large nonpositivity have large uncertainty. 
\end{abstract}

\maketitle

\textit{Introduction---}
One of the most striking differences between quantum and classical theory is that quantum theory does not simultaneously assign definite values to incompatible observables. This peculiarity can be quantified by 
uncertainty principles: for example, given two observables $\hat A$ and $\hat B$, the Heisenberg-Robertson uncertainty relation \cite{Robertson1929} states that
\begin{equation}
    \sigma_{\hat A} \sigma_{\hat B} \geq \frac{1}{2}|\langle[\hat A,\hat B]\rangle|,
\end{equation}
where $\sigma_{\hat A}$ and $\sigma_{\hat B}$ denote the standard deviations of observables $\hat A$ and $\hat B$, respectively. In modern quantum-information theory, one commonly expresses the uncertainty principle in terms of entropies instead of standard deviations \cite{ColesReview}. Let $\hat x$ and $\hat p$ be canonically-conjugate continuous variables and denote by $h(\hat x)$ and $h(\hat p)$ the Shannon differential entropies \cite{shannon1948mathematical} of their respective  outcome-probability distributions.
Then
\begin{eqnarray}\label{eq:BBM}
    h(\hat x) + h(\hat p) \geq \ln(\pi) + 1 \coloneqq \qbm 
\end{eqnarray}
lower bounds the entropic uncertainty associated with $\hat x$ and $\hat p$. The lower bound is saturated on all coherent states \cite{Bialynicki-Birula1975-ks}. 

For finite-dimensional systems, a general entropic uncertainty relation, analogous to Ineq.~\eqref{eq:BBM}, was proven in~\cite{MaassenUffink}:
\begin{equation}\label{eq:maassen_uffink}
    H_1(\hat A) + H_1(\hat B) \geq -\ln(\max_{ab}|U_{ab}|^2) \coloneqq \qmu .
\end{equation}
Here, $H_1(\hat A)=-\sum_a p_a \ln(p_a)$ denotes the Shannon entropy of the outcome-probability distribution $p_a$ for the observables $\hat A$, and similarly for $\hat B$.  $U_{ab}$ is the transition matrix between the eigenbases of the two observables. The bound in Eq.~\eqref{eq:maassen_uffink} is sharp if the two bases are mutually unbiased bases (MUB), in which case $q_{\mathrm{MU}}=\ln d$.\footnote{Two bases $\{ \ket{x_j} \}$ and $\{ \ket{y_k} \}$ are MUB if and only if $|\langle x_j|y_k\rangle| = 1/\sqrt{d}$ for all $j,k$.} Sharper bounds have been proven for non-MUB observables \cite{riccardi_maccone_2026}.

In the analysis of uncertainty, quasiprobability representations of quantum mechanics have played an important role. These representations are tools widely used to study the difference between quantum and classical theory \cite{Wigner1932, Ferrie2011, Arvidsson-Shukur2024}. Quasiprobability distributions represent quantum states with complex-valued measures on a phase (or ontic) space $\Lambda$. They satisfy most of Kolmogorov's axioms for probability distributions, but may take values outside the interval $[0,1]$.  If a state $\hat{\rho}$ has an everywhere positive quasiprobability distribution $Q$, it is said to be ``$Q$ positive''. The set of $Q$-positive states is denoted by $\mathcal E_+^Q$. For quasiprobaility distributions with finite $\Lambda$ we define the $Q$-Shannon entropy $H_1^Q(\hat\rho)$ of any $\hat{\rho} \in \mathcal E_+^Q$:
\begin{equation}
    H_1^Q(\hat\rho)=-
\sum_{\lambda\in\Lambda}Q_\lambda(\hat\rho)\ln(Q_\lambda(\hat\rho)).
\end{equation}
In situations where
$Q_\lambda(\hat\rho)$  reproduces the Born-rule probabilities of the observables $\hat A$ and $\hat B$, $Q_\lambda(\hat\rho)$ is a joint probability distribution for these observables. $H_1^Q(\hat\rho)$ then provides a measure of joint uncertainty of $Q$-positive states, the properties of which we shall compare to those of $H_1(\hat A)+H_1(\hat B)$.

Since $H_1^Q$ is concave, we know, \textit{a priori}, that its minimum is attained on an extremal point of the convex set $\mathcal E_+^Q$. The pure $Q$-positive states are such extremal states, but in general $\mathcal E_+^Q$  has also mixed extremal states. This is the case for the continuous variable Wigner distribution~\cite{brocker1995}, for the Gross-Wigner (GW) distribution $Q^{\mathrm{GW}}$ for $n$ odd dimensional qudits~\cite{Gross2006}, as well as for several Kirkwood-Dirac (KD) distributions $Q^{\mathrm{KD}}$~\cite{Langrenez2024Characterizing,DeBievre2025,Thio2025}. The GW representation has been used extensively in the study of quantum-computational advantages and resource theories \cite{Veitch_2012, Mari12, Veitch_2014, Howard14, Wang_2019, Raussendorf20}. 
Given two arbitrary observables $\hat{A}$ and $\hat{B}$, there is an associated KD representation, which is  uniquely characterized through its associated conditional expectations~\cite{spriet2026a, jordan2026}, and which has been used  in several areas of quantum-information processing. These include metrology \cite{Arvidsson-Shukur2020, lupu2022negative, Hance_2024, Jenne22}, tomography, thermodynamics \cite{YungHalp17, YungHalp18, Razieh19, YungHalp19, Yunger19-2, Gherardini24} and qubit-system computing~\cite{Thio2025}; we refer to~\cite{Lostaglio2023, Arvidsson-Shukur2024} for a review. In settings like these, states with positive quasiprobability distributions are associated with a form of ``classicality'' \cite{Spekkens2008,Veitch_2012, Veitch_2014, Mari12, Arvidsson-Shukur2020,Thio2024,Thio2025}.
Since, intuitively, mixed states are thought of as more uncertain than pure ones, it is natural to ask for which quasiprobability representations the minimum of $H_1^Q$ is reached on a pure $Q$-positive state. In other words, whether  
\begin{equation}\label{eq:graal}
   H_1^Q(\hat \rho)\geq \min_{\hat\rho_+^{\mathrm{pure}}}H_1^Q(\hat\rho^{\mathrm{pure}}_+)
\end{equation}
holds for all $\hat\rho\in \mathcal E_+^Q$.
For the Shannon entropy of the Husimi function \cite{Husimi1940}, known as the Wehrl entropy, this inequality obviously holds since then all states are Husimi positive. 
Wehrl~\cite{WEHRL1979353} conjectured, and Lieb~\cite{Lieb78} proved, that the minimum is reached on coherent states. In~\cite{CARLEN1991}, the coherent states were proven to be the unique minimizers of the Wehrl entropy. For the continuous-variable Wigner distribution, the analogous statement was conjectured~\cite{Hertz_2017, VanHerstraeten2021}, further studied~\cite{VanHerstraeten2023, Cerf2024, VanHerstraeten2025, Dias2023, Qian2024} and recently disproven~\cite{he2026wigner, vanherstraeten2026, liu2026}.

In this Article, we prove, in Theorem~\ref{thm:main1}, the entropic-uncertainty  relation  Eq.~\eqref{eq:graal} for  the GW representation and the KD representations~\cite{Kirkwood33,Dirac45, Lostaglio2023,Arvidsson-Shukur2024} for observables with MUB (see definitions below). We also identify, in each case, the states on which the minimum is reached. Proving Ineq.~\eqref{eq:graal} amounts to showing that, on mixed extremal states, the $Q$-Shannon entropy is larger than or equal to its minimal value on pure $Q$-positive states. However, the set of mixed $Q$-positive extremal states is generally hard to characterize~\cite{Langrenez2024Characterizing,langrenez2024, VanHerstraeten2025,  DeBievre2025, Langrenez2025Convex, Thio2025, vanherstraeten2026_prata}. In our proof, we circumvent the issues with characterizing $\mathcal E_+^Q$ by exploiting the structure of the pure $Q$-positive states for the quasiprobability representations we consider.
Inequality~\eqref{eq:graal} together with marginal properties (see Eq.~\eqref{eq:BornKD} below) of the GW and MUB KD distributions imply that on $ \mathcal{ E}_+^{Q}$
\begin{equation}\label{eq:subadditive2}
  H_1(\hat A)+H_1(\hat B)  \geq H^{Q}_1\geq \min_{ \mathcal{ E}_+^{Q, \text{pure}} }H_1^Q .
\end{equation}
We show, additionally, that 
\begin{equation}\label{eq:graal2}
\min_{  \mathcal{ E}_+^{Q}} \left(H_1(\hat A)+H_1(\hat B)\right)=\min_{  \mathcal{ E}_+^{Q,\text{pure}}} H_1^{Q} .
\end{equation}
Since  $\min_{  \mathcal{ E}_+^{Q, \mathrm{pure}}} H_1^{Q}=\ln d = \qmu $ for the GW and MUB KD distributions, Eq. \eqref{eq:maassen_uffink} is saturated on positive pure states: Minimal $H_1(\hat A)+ H_1(\hat B)$ entropy is equivalent to minimal MUB KD- and GW-Shannon entropy.  Further, to complement Eq. \eqref{eq:subadditive2}, we prove that states with large KD nonpositivity have large uncertainty as measured by Re\'nyi-$1/2$ entropies, and \textit{vice versa}.

\textit{Setup:---}We consider  quasiprobability representations of quantum mechanics, defined using the formalism of frames~\cite{Ferrie2011, Veitch_2012}. Let $\mathcal H$ be a Hilbert space of finite dimension $d$, let $\Lambda$ be a finite set, with cardinality $\left|\Lambda\right|=d^2$. Let $(\hat{S}_\lambda)_{\lambda\in\Lambda}$ be a frame (basis) of $\mathcal L(\mathcal H)$, and let $(\hat{T}_\lambda)_{\lambda\in \Lambda}$ be its unique dual frame, satisfying
\begin{equation}
    \sum_\lambda \hat{S}_\lambda =\mathbb I_{d},\quad \Tr(\hat{T}_\lambda^\dagger \hat{S}_{\lambda'})=\delta_{\lambda, \lambda'}.
\end{equation} 
Given an operator $\hat C$ on $\mathcal H$, we then define 
\begin{equation}\label{eq:defQPRFrame}
    Q_\lambda(\hat C)=\Tr(\hat C S^\dagger_\lambda),\quad \tilde Q_\lambda(\hat C)=\Tr(\hat C \hat{T}^\dagger_\lambda).
\end{equation}
It follows that the maps $Q, \tilde Q:\mathcal L(\mathcal H)\to \mathbb C^\Lambda$ are bijective. 
The pair $(Q,\tilde Q)$, determined completely by the choice of $\Lambda$ and frame $(\hat{S}_\lambda)_\lambda$,  is a quasiprobability representation of quantum mechanics. It associates to each density matrix $\hat \rho$ on $\mathcal H$ a quasiprobability $Q_\lambda(\hat \rho)$ on $\Lambda$, which is a (generally) complex-valued function that satisfies 
\begin{equation}
\sum_\lambda Q_\lambda(\hat \rho) = 1.
\end{equation}
Moreover, to each observable $\hat C=\hat C^\dagger \in\mathcal L(\mathcal H)$, its symbol $\tilde Q_\lambda(\hat C)$ (or classical observable) provides a quasiprobabilistic expectation value:
\begin{equation}\label{eq:overlap}
    \Tr(\hat \rho \hat C)=\sum_\lambda Q_\lambda(\hat\rho)\overline{\tilde Q_\lambda(\hat C)} .
\end{equation}
Given a state $\hat\rho$ on $\Hcal$, we define its $Q$-nonpositivity as
\begin{equation}
\label{Eq:Neg}
\mathcal N(\hat\rho)=\sum_\lambda |Q_\lambda|-1.
\end{equation}
In general,  $\mathcal N(\hat\rho)\geq 0$, and $\mathcal N(\hat\rho)=0$ if and only if $Q_{\lambda}(\hat \rho)\geq 0$ for all $\lambda\in \Lambda$.  
In the latter case, we call $\hat{\rho}$  $Q$-positive. We denote by $\mathcal E_+^Q$ the convex set of all $Q$-positive states. The $Q$ mana, 
\begin{equation}
\label{Eq:Mana}
    \mathcal M(\hat\rho)= \ln \left[ \mathcal N(\hat\rho) + 1 \right],
\end{equation}
forms an additive resource monotone of $Q$ nonpositivity \cite{Veitch_2014,Thio2025}.

 In addition, given a quasiprobability representation $(Q,\tilde Q)$, we define the associated Margenau-Hill (MH) distribution $Q^{\mathrm{MH}}(\hat\rho)$ as
\begin{equation}
Q^{\mathrm{MH}}(\hat\rho)=\mathrm{Re} Q(\hat\rho), \text{  where  } \sum_{\lambda} Q^{\mathrm{MH}}_\lambda(\hat\rho)=1.
\end{equation}
Clearly, the MH distribution of a real-valued distribution is just the distribution itself. 
We designate the set of states with positive MH distribution by $\mathcal E^{\mathrm{MH}}_+$. One has that $\mathcal E_+^Q\subset \mathcal E_+^{\mathrm{MH}}$.
We write $H_1^{\mathrm{MH}}(\hat\rho)$ for the corresponding MH-Shannon entropy of $\hat\rho\in \mathcal E_+^{\mathrm{MH}}$:
\begin{equation}
   H_1^{\mathrm{MH}}(\hat\rho) =-\sum_\lambda  Q^{\mathrm{MH}}_\lambda(\hat\rho) \ln Q^{\mathrm{MH}}_\lambda(\hat\rho).
\end{equation}

\begin{definition} 
A quasi-probability representation is said to be self-dual if there exists a $\kappa:\lambda\in\Lambda\to\kappa_\lambda\in\R^+$  such that 
\begin{equation}
\hat{T}_\lambda= \kappa_\lambda^{-1} \hat{S}_\lambda.
\end{equation}
It is said to be strongly self-dual if $\kappa_\lambda=\kappa\in\R^+$ is independent of $\lambda$. 
\end{definition}
One family of self-dual quasiprobability representations are the KD representations $(Q^{\mathrm{KD}}, \tilde Q^{\mathrm{KD}})$ \cite{Kirkwood33, Dirac45, Lostaglio2023,Arvidsson-Shukur2024}. They are determined by two complete sets of commuting observables $\hat A=(\hat A_1\dots \hat A_n)$, $\hat B=(\hat B_1,\dots, \hat B_m)$  with 
\begin{equation}\label{eq:KDdef}
\hat{S}_{a,b}=\hat\Pi_a\hat\Pi_b,\quad  \hat{T}_{a,b}=\kappa_{a,b}^{-1}  \hat{S}_{a,b}, \quad \kappa_{a,b}=|\langle a|b\rangle|^2 .
\end{equation}
Here $a=(a_1,\dots, a_n)\in\sigma(\hat A)$ is a vector in the joint spectrum of $\hat{A}$ and similarly for $b\in \sigma( \hat B)$; and $\Pi_a=|a\rangle\langle a|$, $\Pi_b=|b\rangle\langle b|$. Hence, $\Lambda=\sigma(\hat A)\times \sigma(\hat B)$ and $\lambda=(a,b)$. Informational completeness of the KD framework requires that $\kappa_{a,b} \neq 0$ for all $a,b$ \cite{Arvidsson-Shukur2024}. If the bases of $\hat{A}$ and $\hat{B}$ are MUB, the KD representation is strongly self-dual, with $\kappa_{a,b}=d^{-1}$. The KD distributions $Q_{a,b}^{\mathrm{KD}}(\hat\rho)$ are thus defined by using the KD frames [Eq. \eqref{eq:KDdef}] in Eq. \eqref{eq:defQPRFrame}.  The KD representations are Born compatible:
\begin{equation}
\label{eq:BornKD}
    \langle a|\hat\rho|a\rangle=\sum_{b\in\sigma( \hat B)} Q_{a,b}^{\mathrm{KD}}(\hat\rho), \quad \langle b|\hat\rho|b\rangle=\sum_{a\in\sigma(\hat A)} Q_{a,b}^{\mathrm{KD}}(\hat\rho).
\end{equation}
The KD representations correspond to left ordering, in the sense that, for all $n,m\in\mathbb N$, $\tilde Q_{a,b}(\hat A^n \hat B^m)=a^n b^m$: the operator $\hat A^n\hat B^m$ associated to the classical monomial $a^nb^m$ puts powers of $\hat A$ to the left of those of $\hat B$. We note that $(Q^{\mathrm{KD}}, \tilde Q^{\mathrm{KD}})$ depend only indirectly on the observables $\hat{A}$ and $\hat{B}$ via their transition matrix $U_{ab}=\langle a|b\rangle$.

Also the GW representation $(Q^{\mathrm{GW}}, \tilde Q^{\mathrm{GW}})$~\cite{Gross2006} is strongly self-dual. It is associated to a discrete position and momentum, and corresponds to a symmetric ordering, in analogy to the original Wigner-Weyl representation. To define the GW representation, consider a Hilbert space of odd dimensionality $d$.   Define generalized Pauli operators $\hat{X}$ and $\hat{Z}$ such that $\hat{X} \ket{ a } = \ket{a + 1 \mod d}$ and $\hat{Z} \ket{ a } = \omega^a \ket{a}$, respectively. Here, $\omega = e^{2 \pi i / d}$. Further, let $\hat{D}_{a,b} =  \omega^{a b (d+1)/2} \hat{X}^a \hat{Z}^b$, where $a,b$ take values from $1$ to $d$, modulo $d$.   The GW phase-space operators are then
\begin{equation}
\label{eq:GWFrame1}
    \hat{S}_{a,b} =  \hat{D}_{a,b} \hat{S}_{0,0} \hat{D}_{a,b}^{\dagger} , \text{  where  } \hat{S}_{0,0} = \frac{1}{d^2} 
 \sum_{a,b} \hat{D}_{a,b} .
\end{equation}
Hence, $\Lambda = \mathbb Z_d\times \mathbb Z_d$.
The dual frame is given by
\begin{equation}
\label{eq:GWFrame2}
    \hat{T}_{a,b} = d \hat{S}_{a,b} ,
\end{equation}
such that $\kappa=d^{-1}$.  The GW distribution $Q_{a,b}^{\mathrm{GW}}(\hat\rho)$ is thus defined by using the GW frames [Eqs. \eqref{eq:GWFrame1} and \eqref{eq:GWFrame2}] in Eq. \eqref{eq:defQPRFrame}. We write $|b\rangle$, with $b$ an integer modulo $d$, for the eigenbasis of $\hat{X}$: $\hat{X}|b\rangle=\omega^{-b}|b\rangle$. The transition matrix $\langle a|b\rangle=\omega^{ab} / \sqrt{d}$ is the discrete-Fourier-transform matrix.   Like the KD representations, the GW representation is Born compatible. Moreover, like the GW representation, the KD representation can be defined with respect to the Fourier transform $U_{a,b} =\omega^{ab} / \sqrt{d}$.

Both the KD and GW representations are readily extended to systems of $n$ qudits with $d \rightarrow d^n$~\cite{Gross2006,DeBievre2025, spriet2025}. 
Of particular use in the study of quantum computation is the fact that the pure positive GW states are the stabilizer states \cite{Gross2006, Veitch_2012, Mari12, Veitch_2014}. However, mixed extremal states exist outside the stabilizer polytope \cite{Gross2006}. Thus, the set of ``classical'' states is more complicated than just the stabilizer polytope. In the quasiprobabilistic study of quantum computing, a major shortcoming of the GW algorithm is its limitation to odd $d$; the most common quantum-information systems are made up of $n$ qubits. In such scenarios, the MUB KD distribution with $\hat{A} = \hat Z$ and $\hat{B}= \hat X$ is a useful alternative \cite{Thio2025}.  The 
pure positive states of these KD distributions are the Calderbank-Shor-Steane stabilizer states, which are of particular importance in quantum error correction \cite{DeBievre2025, Thio2025}. As with the GW distribution, mixed extremal states exist outside the Calderbank-Shor-Steane stabilizer polytope \cite{Thio2025}. To summarise, the characterizations of all extremal states of both families of distributions are not straightforward. Consequently, other techniques are needed to prove [Ineq. \eqref{eq:graal}] that the $Q$-Shannon entropy on mixed extremal states is larger than or equal to its minimal value on
pure $Q$-positive states.

\textit{The main result:---} 
We can now state our main result as follows:
\begin{Theorem}\label{thm:main1}
Let $(Q,\tilde Q)$ either be a MUB KD representation, or a GW representation of $n$ odd-dimensional qudits. Then the following hold:\\
(i) For all $\hat \rho \in  \mathcal E_+^{\mathrm{MH}}$,\begin{equation}
\label{eq:ineqstringKDGW}
 H_1(\hat A)+H_1(\hat B)\geq H^{\mathrm{MH}}_1(\hat\rho)
\geq -\ln\left(\Tr(\hat\rho^2)\right)+\ln d \geq \ln d;
\end{equation}
(ii) $H^{\mathrm{MH}}_1(\hat\rho)=\ln d$ if and only if $\hat\rho\in\mathcal E^{Q,\mathrm{pure}}_+$;\\
(iii) If $(Q,\tilde Q)$ is a MUB KD representation, then all  $\hat{\rho} \in \mathcal E_+^{\mathrm{KD, pure}}$ satisfy  $H_1(\hat A)+H_1(\hat B)=\ln d$;
\\
(iv) If $(Q,\tilde Q)$ is a GW representation, then
\begin{equation}
    \min_{\mathcal E_+^{\mathrm{GW, pure}}}(H_1(\hat A)+H_1(\hat B))=\ln d.
\end{equation}
\end{Theorem}
Parts (i) and (ii) together imply Ineq.~\eqref{eq:graal}, both for the MUB KD representations and for the GW representations. In addition, (i) implies that, if $\hat\rho\in \mathcal E_+^{Q,\mathrm{ extr.}}$, the set of extremal states of the convex set $\mathcal E_+^{Q}$,  but is not pure, then $H_1^{\mathrm{MH}}(\hat\rho)>\ln d$.  For the KD distributions, the result is in fact stronger than what is indicated in Ineq.~\eqref{eq:graal}, since the bound is shown to hold for all MH-positive states, and not only for the KD-positive ones; recall that $\mathcal E_+^{\mathrm{MH}} \supseteq \mathcal E_+^{\mathrm{KD}}$. Part~(iii)  states that all KD-positive pure states saturate Ineq. \eqref{eq:maassen_uffink}: $H_1(\hat A)+H_1(\hat B)=\ln d$. This implies Eq.~\eqref{eq:graal2}.
As indicated by (iv), the situation is different for the GW distribution. Indeed, there exist GW-positive pure states for which $H_1(\hat A)+H_1(\hat B)=2\ln d$. 

It is natural to ask whether the lower bound on $H_1^{\textrm{MH}}(\hat\rho)$ is also saturated at fixed purity. we investigate this question in Appendix~\ref{s:fixedpurity}.

The following proposition, which is of interest by itself, provides the first ingredient of the proof of Theorem~\ref{thm:main1}.
\begin{Proposition}\label{prop:H1MHselfdualbound}
Let $(Q,\tilde Q)$ be a self-dual quasiprobability representation of quantum mechanics.
 Let $\hat\rho\in \mathcal E_+^{\mathrm{MH}}$. Then
\begin{eqnarray}\label{eq:H1MHselfdualbound}
H^{\mathrm{MH}}_1(\hat\rho)
&\geq& -\ln\left(\Tr(\hat\rho^2)\right)-\ln \|\kappa\|_\infty \geq \ln \|\kappa\|^{-1}_\infty.
\end{eqnarray}
If $\hat\rho\in \mathcal E_+^{\mathrm{MH}}$ saturates the lower bound, so that
\begin{equation}
H^{\mathrm{MH}}_1(\hat\rho)= \ln\|\kappa\|_\infty^{-1},
\end{equation} 
then $\hat\rho$ is pure  and $Q$-positive. 
\end{Proposition}
\begin{proof}
To prove the first statement, we use Jensen's inequality:
\begin{eqnarray}\label{eq:H1MHselfdualboundbis}
H^{\mathrm{MH}}_1(\hat\rho)&=&-\sum_\lambda  Q^{\mathrm{MH}}_\lambda(\hat\rho) \ln Q^{\mathrm{MH}}_\lambda(\hat\rho).\nonumber\\
&\geq& -\ln\left(\sum_\lambda (Q^{\mathrm{MH}}_\lambda(\hat\rho))^2\right).
\end{eqnarray} 
For any $\hat\rho$,
\begin{eqnarray}
\sum_\lambda |Q_\lambda^{\mathrm{MH}}(\hat\rho)|^2&\leq& \sum_\lambda |Q_\lambda(\hat\rho)|^2=
\sum_\lambda \kappa_\lambda Q_\lambda(\hat\rho)\overline{\tilde Q_\lambda(\hat\rho)}\nonumber\\
&\leq& \|\kappa\|_\infty \Tr(\hat\rho^2).
\end{eqnarray}
Here, we used Eq.~\eqref{eq:overlap} and 
\begin{equation}
    \mid Q_\lambda (\hat \rho)\mid = \sqrt{\kappa_\lambda}\sqrt{Q_\lambda(\hat\rho)\overline{\tilde Q_\lambda(\hat\rho)}}.
\end{equation} Hence, if $\hat \rho$ is MH positive, then 
\begin{equation}
\sum_\lambda Q_\lambda^{\mathrm{MH}}(\hat\rho)^2\leq \|\kappa\|_\infty \Tr(\hat\rho^2).
\end{equation}
Using this in Eq.~\eqref{eq:H1MHselfdualboundbis}, Eq.~\eqref{eq:H1MHselfdualbound} follows. 

To prove the second statement, note that, if $\hat \rho$ is mixed, then it cannot saturate the lower bound in Eq.~\eqref{eq:H1MHselfdualbound}. Next, suppose that $\hat \rho$ is pure and belongs to  $\mathcal E_+^{\mathrm{MH}}$ but not to $\mathcal E_+^{\mathrm{KD}}$. 
Then 
\begin{equation}
\sum_\lambda (Q_\lambda^{\mathrm{MH}}(\hat \rho))^2< \sum_\lambda |Q_\lambda(\hat \rho)|^2\leq \|\kappa\|_\infty,
\end{equation}
where the first inequality is strict. Consequently Eq.~\eqref{eq:H1MHselfdualboundbis} implies that
$
H_1^{\mathrm{MH}}(\hat\rho)> \ln \|\kappa\|^{-1}_\infty.
$
Hence $\hat\rho$ does not saturate the lower bound in Eq.~\eqref{eq:H1MHselfdualbound}. This completes the proof. 
\end{proof}
The converse of the last statement in  Proposition~\ref{prop:H1MHselfdualbound} is in general false. We give an example in Appendix~\ref{s:nonMUB} of a non-MUB KD quasiprobability representation for which the lower bound in Eq~\eqref{eq:H1MHselfdualbound} is not reached on any of the pure KD-positive states. 
On the other hand, Theorem~\ref{thm:main1} shows that, for MUB KD representations, and the GW representations of $n$ odd-dimensional qudits, the converse does indeed hold.   We now proceed to prove Theorem \ref{thm:main1}.

\begin{proof}
(i) The first inequality follows from the subadditivity of the Shannon entropy, together with the fact that the KD and GW quasiprobabilities are Born compatible for $\hat A$ and $\hat B$. The second inequality follows from Proposition~\ref{prop:H1MHselfdualbound} and the fact that $\|\kappa\|_\infty=d^{-1}$.
(ii) The direct implication follows from  Proposition~\ref{prop:H1MHselfdualbound}. The converse implication follows from the known structure of the pure positive states. We first consider MUB KD representations. In~\cite{DeBievre2023, Xu2024} the following properties of pure  MUB KD-positive states were proven. Let $S_{\hat A}=\{a\in\sigma(\hat A) \, | \, \langle a|\psi\rangle \not=0\}$, and define $n_{\hat A}= | S_{\hat A}|$, and similarly for $S_{\hat B}, n_{\hat B}$. Then, for all $a\in S_{\hat A}$, $|\langle a|\psi\rangle|
=n_{\hat A}^{-1/2}$; similarly for $\hat B$. In addition $n_{\hat A} n_{\hat B}=d$. Finally, $Q_{a,b}(\psi)=d^{-1}$  if $(a,b)\in S_{\hat A}\times S_{\hat B}$ and vanishes otherwise. The converse implication in (ii) now follows immediately for the MUB KD representations. Part~(iii) follows from the preceding properties of the KD-positive pure states. 
It is proven in~\cite{Gross2006} that also the GW distributions of pure GW-positive states are constant on their support, which has cardinality $d$. This proves (ii) and (iv).
\end{proof}

\textit{Upper bounding KD Nonpositivity by Uncertainty:---}Above, we lower bounded the $Q$-Shannon entropy of $Q$-positive states. Here we look at a complementary question: Does a large $Q$-negativity imply a large uncertainty? We answer this question affirmatively for MUB KD distributions. For MUB KD distributions, the nonpositivity [Eq. \eqref{Eq:Neg}] is bounded:
\begin{equation}\label{eq:Nupperbd}
\mathcal N(\hat\rho)\leq-1+\sum_{a,b}|\braket{a}{b}|\,\sqrt{p_a q_b},
\end{equation}
where $p_a=\langle a|\hat\rho|a\rangle$ and $q_b=\langle b|\hat\rho|b\rangle$. Here, we used the Cauchy-Schwarz inequality to bound
\begin{equation}\label{eq:qjkbound}
|\langle a|\hat\rho|b\rangle|=|\langle a|\hat\rho^{1/2}\hat\rho^{1/2}|b\rangle|\leq p_a^{1/2}q_b^{1/2}.
\end{equation} 
Since all $ |\bra{a}\ket{b}| = 1 / \sqrt{d}$, we have that
\begin{equation}
\label{Eq:UpBound}
\mathcal N(\hat\rho)\leq-1+ \frac{1}{\sqrt{d}} \left( \sum_{j}\,\sqrt{p_j } \right) \left( \sum_{k}\,\sqrt{q_k } \right).
\end{equation}
 By rearranging Eq. \eqref{Eq:UpBound} and taking a logarithm, we bound the KD mana [Eq. \eqref{Eq:Mana}]:
\begin{equation}
    \mathcal M(\hat\rho)\leq - \ln(\sqrt{d}) + \frac{1}{2} \left[ H_{\frac{1}{2}}  (\hat{A})  +  H_{\frac{1}{2}}(\hat{B}) \right] \leq \ln(\sqrt{d}).
\end{equation}
Here,
\begin{equation}
    H_{\alpha}(\hat{A} ) = \frac{1}{1-\alpha} \ln \left( \sum_j p_j^{\alpha}\right) ,
\end{equation}
is the Re\'nyi-$\alpha$ entropy of the observable $\hat{A}$, and similarly for $H_{\alpha}(\hat{B} )$. Since pure states saturate Ineq. \eqref{eq:Nupperbd}, we have that $\mathcal M(\ket{\psi}\bra{\psi})$ is affinely related to uncertainty as measured by $H_{\frac{1}{2}}  (\hat{A})  +  H_{\frac{1}{2}}(\hat{B})$.
States with large mana, therefore, necessarily have large uncertainty. Hence, KD nonpositivity is a source of quantum uncertainty for arbitrary observables. Conversely, low uncertainty implies low mana. KD mana is a resource for quantum advantages in several settings \cite{Thio2025, Arvidsson-Shukur2020, lupu2022negative}; low-uncertainty states have limited resourcefulness in such settings.


\textit{Discussion:---}Given a self-dual quasiprobability representation $(Q,\tilde Q)$ of quantum mechanics, we have obtained a Shannon-entropic uncertainty relation for the states $\hat\rho$ with positive quasi-probability distribution $Q(\hat\rho)$ (Proposition~\ref{prop:H1MHselfdualbound}).
We have shown that, for Kirkwood-Dirac representations with mutually unbiased bases and for Gross-Wigner representations of $n$ odd-dimensional qudits, this bound is sharp and saturated only on the pure positive states of the quasiprobability representations considered (Theorem \ref{thm:main1}).

These results stand in contrast to the situation for the continuous-variable Wigner function $W$ (elaborated on in Appendix \ref{app:WignerShannon}). Indeed, when applied to $W$, the reasoning used to prove Ineq.~\eqref{eq:H1MHselfdualbound}, together with subadditivity and Born compatibility  leads to 
\begin{equation}
    H_1(\hat x)+ H_1(\hat p)\geq H_1^{\mathrm{W}}(\hat \rho)> -\ln\left(\Tr\hat\rho^2\right)+ \ln \pi.
\end{equation}
However, since continuous variable Wigner functions cannot be constant on their support, the second inequality is now sharp, contrary to what happens in the discrete variable quasiprobability representations considered in this Article. In addition, Ineq.~\eqref{eq:BBM}, together with the Hudson Theorem \cite{Hudson1974}, imply that, for the continuous variable Wigner function 
\begin{equation}
    \min_{|\psi\rangle\langle \psi|\in \mathcal E_+^{\mathrm{W}}} H^{\mathrm{W}}(|\psi\rangle\langle \psi|)=\ln\pi+1.
\end{equation}
For Wigner functions, Ineq.~\eqref{eq:graal} therefore reads: for all $\hat \rho\in \mathcal E^{\mathrm{W}}_+,$
\begin{equation}\label{eq:graal_W}
H_1^{\mathrm{W}}(\hat\rho)\geq \ln\pi+1.
\end{equation}
This is precisely the Wigner-Shannon conjecture proposed in~\cite{Hertz_2017, VanHerstraeten2021} and that was recently disproven~\cite{vanherstraeten2026, liu2026}. In other words, the analogs of (ii) and (iv) of Theorem~\ref{thm:main1} do not hold for the continuous variable Wigner function. One concludes that the notion of uncertainty captured by the Wigner-Shannon entropy is different from the one captured by the Shannon entropy sum $H(\hat x)+H(\hat p)$.
 
The relation between the KD-Shannon entropy and the entropy $H_1(\hat A)+H_1(\hat B)$ is similarly  less straightforward for non-MUB KD distributions compared to MUB ones. We provide examples  in Appendix~\ref{s:nonMUB}, where we conjecture that, for non-MUB KD distributions, with probability $1$, Ineq.~\eqref{eq:graal2} is violated. while Ineq.~\eqref{eq:graal} holds. We also provide a non-MUB KD counterexample to  Ineq.~\eqref{eq:graal}.

\medskip

\textit{Acknowledgements---}The analytical results in this manuscript were all produced without the use of AI. Some of the numerical examples given in the appendices were constructed in collaboration with AI. JJT was supported by the Cambridge Trust.  This work was supported in part by the Agence Nationale de la Recherche under grant ANR-11-LABX0007-01 (Labex CEMPI), by the Nord-Pas de Calais Regional Council and the European Regional Development Fund through the Contrat de Projets \'Etat-R\'egion (CPER), and by the CNRS through the MITI interdisciplinary programs.

\medskip

\section*{Appendices}

\appendix
\crefalias{section}{appendix}

\section{Minimizing the KD-Shannon entropy at fixed purity for MUB KD and GW representations}\label{s:fixedpurity}
In this appendix, we consider strongly self-dual quasi-probability representations for which 
\begin{equation}\label{eq:B1}
    \tilde Q_{\lambda}(\mathbb{I})=1, \forall \lambda\in\Lambda.
\end{equation}
In other words, we assume that the quantum observable $\hat C=\mathbb{I}$ is represented by the constant classical observable, taking the value $1$ everywhere.   This implies that $Q_\lambda(d^{-1}\mathbb{I})=\|\kappa\|_{\infty}d^{-1}$
is the uniform probability distribution on $\Lambda$ and hence that $\kappa_{\lambda}=d^{-1}$. The KD MUB representations as well as the GW representations all have this property. Non-MUB KD distributions do not: for them, the maximally mixed state is not represented by the uniform probability distribution on $\Lambda$, even though Eq.~\eqref{eq:B1} holds. 

Given such a quasiprobability representation, Ineq.~\eqref{eq:ineqstringKDGW} holds in the form
\begin{equation}\label{eq:H1puritybound}
H_1(\hat\rho)\geq -\ln(\Tr(\hat\rho^2))+\ln d.
\end{equation} 
On can therefore ask if, at fixed purity, this inequality is saturated. In other words, we are looking for the minimum of $H_1(\hat \rho)$ over all $Q$-positive states with a fixed purity $\Tr(\hat\rho^2)=\tau$, $\frac1d\leq \tau\leq 1$. The question is if this minimum saturates the lower bound in Eq.~\eqref{eq:H1puritybound}. 

Our first result on this question is summed up in the following lemma.
\begin{Lemma}\label{lem:puritysharpbound}
Let $(Q, \tilde Q)$ be a strongly self-dual quasiprobability representation on $\Lambda$ for which Eq.~\eqref{eq:B1} holds. Suppose that $\hat\rho\in \mathcal E_+^{\mathrm{MH}}$ saturates  Ineq.~\eqref{eq:H1puritybound}. Then $\hat\rho\in \mathcal E_+^Q$, and $Q_{\lambda}(\hat \rho)$ is constant on its support. In addition, writing $D$ for the size of this support,  one has $d\leq D \leq d^2$ and $d^{-1}\leq \Tr(\hat\rho^2)=d{D}^{-1}\leq 1$.
\end{Lemma}
The Lemma asserts that Ineq.~\eqref{eq:H1MHselfdualbound} can be saturated only for a finite number of very specific values of the purity and, in addition, that this minimum occurs necessarily on a $Q$-positive state. For which values  of $D$ the lower bound is effectively reached is not generally obvious. Below we treat some specific cases where this question can be answered. 
\begin{proof} 
Using Eq.~\eqref{eq:overlap}, we find that, for any $\hat \rho$,
\begin{equation}\label{eq:B3}
    \Tr(\hat \rho^2)=d\sum_{\lambda} Q_{\lambda}(\hat\rho)\overline{Q_{\lambda}(\hat \rho)}=d\sum_{\lambda} |Q_{\lambda}(\hat\rho)|^2.
\end{equation}
Hence, if Ineq.~\eqref{eq:ineqstringKDGW} is saturated for some $\hat \rho\in \mathcal E_+^{\mathrm{MH}}$, then 
\begin{eqnarray}
    H_1^{\mathrm{MH}}(\hat \rho)&=&-\ln\left(d^{-1}\Tr\hat\rho^2\right)\\
    &=&-\ln\left(\sum_{\lambda} |Q_{\lambda}(\hat\rho)|^2\right)\label{eq:B5}\\
    &\leq &-\ln\left(\sum_{\lambda} Q_{\lambda}^{\mathrm{MH}}(\hat\rho)^2\right).
\end{eqnarray}
Hence, Ineq.~\eqref{eq:H1MHselfdualboundbis} implies that, for all $\hat\rho\in \mathcal E_+^{\mathrm{MH}}$,
\begin{equation}
    H_1^{\mathrm{MH}}(\hat \rho)=-\ln\left(\sum_{\lambda} Q_{\lambda}^{\mathrm{MH}}(\hat\rho)^2\right).
\end{equation}
Since the right hand side of this equation is the Renyi 2-entropy, it follows that $Q_{\lambda}^{\mathrm{MH}}(\hat \rho)$ is constant on its support. We write $D$ for the size of this support, so that Eq.~\eqref{eq:B3} reads
\begin{equation}
\Tr(\hat\rho^2)=\frac{d}{D}.
\end{equation}
It then follows from $d^{-1}\leq \Tr(\hat\rho)^2 \leq 1$  that $D$ is an integer $d\leq D\leq d^2$. In addition, it follows from  Eq.~\eqref{eq:B5} that $\mathrm{Im} Q_{a,b}(\hat\rho)=0$, so that in fact, $\hat\rho\in \mathcal E_+^Q$.
\end{proof}
The lemma does not identify the values of $d\leq D\leq d^2$ for which the bound in Ineq.~\eqref{eq:H1puritybound} is effectively saturated. We start by showing this is always the case when $D=\delta d$. for all $1\leq \delta \leq d$ for MUB KD. 
To show this, we consider 
\begin{equation}\label{eq:B9}
    \hat\rho=\sum_{a} \alpha_{a}|a\rangle\langle a|,\quad d^{-1}\leq \Tr(\hat\rho^2)=\sum_a\alpha_a^2\leq 1,
\end{equation}
for which one finds
\begin{equation}
    Q^{\mathrm{KD}}_{a, b}=\alpha_{a} \frac1d,
\end{equation}
and
\begin{equation}
    H_1(\hat\rho)=-\sum_{a,b} \ln(\alpha_{a}\frac1d) \alpha_{a}\frac1d=-\sum_a\alpha_a\ln\alpha_a+\ln d.
\end{equation}
The question is if one can adjust the $\alpha_a$ so that this equals the right-hand side of Ineq.~\eqref{eq:ineqstringKDGW}. But this means that one needs 
\begin{equation}
    -\sum_a\alpha_a\ln \alpha_a= -\ln\left(\sum_a \alpha_a^2\right).
\end{equation}
This can only happen  if the $\alpha_a$ are all identical on their support. So $\alpha_a=1/\delta$ for some integer $1\leq \delta\leq d$ and has only $\delta$ non-zero values. It then follows that 
\begin{equation}\label{eq:B13}
\tau=1/\delta=\frac{d}{D},
\end{equation}
where $D=\delta d$ is the size of the support of $Q(\hat \rho)$. The same argument proves the result also for GW representations since the basis states are also GW positive. 

In conclusion, given any MUB KD representation, or GW representation, Ineq~\eqref{eq:ineqstringKDGW} is saturated whenever $\tau=d/D$ for any integer $D=\delta d$, with $1\leq \delta\leq d$.

Whether these are the only possible $D$-values for which saturation occurs is not clear in general. We do have the following result, nevertheless.
\begin{Lemma}
Consider a MUB KD representation for which 
\begin{equation}\label{eq:Esimple}
    \mathcal E_+^{\mathrm{KD}}=\mathrm{conv} \{|a\rangle\langle a|, |b\rangle\langle b|\mid a,b\}.
\end{equation}
Then Ineq.~\eqref{eq:H1puritybound} is saturated at fixed purity if and only if  $\Tr\hat\rho^2=\frac{d}{D}$, with $D=\delta d$, $1\leq \delta \leq d$.
\end{Lemma}
\begin{proof}
    The if part was proven above. For the only if part, we note that any $\hat\rho\in\mathcal E_+^{\mathrm{KD}}$ can be written, with $0\leq\alpha_a, \beta_b\leq 1$, 
\begin{equation}
    \hat\rho=\hat\rho_A+\hat\rho_B,
\end{equation}
where
\begin{equation}
    \hat\rho_A=
    \sum_a \alpha_a |a\rangle\langle a|,\hat \rho_B=\sum_b\beta_b|b\rangle\langle b|, \sum_a\alpha_a+\sum_b\beta_b=1.
\end{equation}
Note that this decomposition is not unique since 
\begin{equation}
    \sum_a |a\rangle\langle a|=\mathbb{I}=\sum_b |b\rangle\langle b|,
\end{equation}
but this nonuniqueness will not affect the argument below. 
Consequently, one has
\begin{equation}
Q_{a,b}^{\mathrm{KD}}=\frac1{d}\left(\alpha_a+\beta_b\right).
\end{equation}
We assume that $\hat\rho_B\not=0$ since otherwise the reasoning in Eq.~\eqref{eq:B9}-\eqref{eq:B13} yields the conclusion. There exists therefore $1\leq b'\leq d$ so that $\beta_{b'}\not=0$. We know furthermore from Lemma~\ref{lem:puritysharpbound} that $Q_{a,b}$ is constant on its support. Since, for all $1\leq a\leq d$, $(a,b')$ belongs to the support of $Q^{\mathrm{KD}}$, this implies that 
\begin{equation}
\alpha_a+\beta_{b'}=\alpha_{a'}+\beta_{b'},\quad \forall 1\leq a,a'\leq d.
\end{equation}
Hence there exists $0\leq \alpha\leq 1$ so that $\alpha_a=\alpha$ for all $a$ and therefore
\begin{equation}
    \hat\rho=\sum_b (\alpha +\beta_b)|b\rangle\langle b|,
\end{equation}
which allows to conclude by applying the argument in Eq.~\eqref{eq:B9}-\eqref{eq:B13} with the roles of $a$ and $b$ interchanged.
\end{proof}
The condition Eq.~\eqref{eq:Esimple} is satisfied for the KD distributions arising from the discrete Fourier transform on qudits, with $d$ a prime number, as proven in~\cite{Langrenez2024Characterizing}.


\section{Non-MUB KD representations}\label{s:nonMUB}
We first consider the KD quasiprobability representation of quantum mechanics  with the non-MUB transition matrix
\begin{equation}\label{eq:Upet}
U=
\frac13\begin{pmatrix}
-1&2&2\\
2&-1&2\\
2&2&-1
\end{pmatrix},
\end{equation}
first considered in~\cite{Langrenez2024Characterizing}.
The bound from Ineq. \eqref{eq:maassen_uffink} is then
$$
H_1(A)+H_1(B)\geq q_{\mathrm{MU}}=2\ln(3/2)\approx 0.81.
$$
This bound is not sharp. We numerically determined the minimum of $H_1(A)+H_1(B)$ to be $0.9348$.
The full set $\mathcal E_+^{\mathrm{KD}}$ of 9 pure KD-positive states for this case was identified in~\cite{Langrenez2024Characterizing}. Beyond the six basis states, there are three further KD-positive  states:
\[
\cket{\varphi_3} = \frac{\cket{a_1}-\cket{a_2}}{\sqrt{2}},\
\cket{\varphi_2} = \frac{\cket{a_3}-\cket{a_1}}{\sqrt{2}}, \
\cket{\varphi_1} = \frac{\cket{a_2}-\cket{a_3}}{\sqrt{2}}.
\]
Their associated KD distributions are
\[
Q^{\mathrm{KD}}(\varphi_3) = \frac{1}{6}\begin{pmatrix} 1 & 2 & 0 \\ 2 & 1 & 0 \\ 0 & 0 & 0 \end{pmatrix}, \ Q^{\mathrm{KD}}(\varphi_2) = \frac{1}{6}\begin{pmatrix} 1 & 0 & 2 \\ 0 & 0 & 0 \\ 2 & 0 & 1 \end{pmatrix}, 
\] and 
\[
Q^{\mathrm{KD}}(\varphi_1) = \frac{1}{6}\begin{pmatrix} 0 & 0 & 0 \\ 0 & 1 & 2 \\ 0 & 2 & 1 \end{pmatrix}.
\]
respectively. Here $Q^{\mathrm{KD}}(\varphi_i):=Q^{\mathrm{KD}}(|\varphi_i\rangle\langle \varphi_i|)$.  Unlike KD-positive pure states for MUB distributions, these states are not constant on their supports. It is in addition established in~\cite{Langrenez2024Characterizing} that, for this distribution, 
\begin{equation}
\mathrm{conv}(\mathcal E_+^{\mathrm{KD, pure}})\subsetneq \mathcal E_+^{\mathrm{KD}},
\end{equation}
so that there exist mixed extremal KD-positive states.

On all basis states $|a_i\rangle, |b_j\rangle$, $i,j=1,2,3$,
$$
H_1(\hat A)+H_1(\hat B)=\frac19(\ln 9+8\ln(\frac{9}{4})\approx 0.965,
$$
and 
$$
H^{\mathrm{KD}}_1=\frac19(\ln 9+8\ln(\frac{9}{4})\approx 0.965.
$$
The fact that these entropies do not depend on the basis state chosen is a result here of the special structure of $U$: all its rows and columns are, up to permutation, identical.
On the other hand, the fact that, on each basis state,  $H_1(\hat A)+H_1(\hat B)=H_1^{\mathrm{KD}}$ is a general feature of the KD representations since for those states, $Q^{\mathrm{KD}}$ is a product distribution so that equality follows from additivity of the entropy for independent probabilities. 

The three remaining KD-positive pure states have uniform marginal distributions supported on two points, for both $\hat A$ and $\hat B$, so that
$$
H_1(\hat A)=H_1(\hat B)=\ln 2,\quad H_1(\hat A)+H_1(\hat B)=2\ln2 \approx 1.38.
$$
In addition, for these three states
 $$
 H^{\mathrm{KD}}(\varphi_i)=\ln(54)^{1/3}\approx 1.329> 2 \ln(3/2)=q_{MU}. 
 $$
 Consequently, on these three states
 \begin{equation}
     H_1(\hat A)+H_1(\hat B)>H^{\mathrm{KD}}_1(\varphi_i)> q_{\mathrm{MU}}.
 \end{equation}
Note that, therefore, the KD-positive pure states do not all have the same KD-Shannon entropy in this example non-MUB KD distribution, nor the same value for $H_1(\hat A)+H_1(\hat B)$, unlike the situation for MUB KD distributions.  Also, in this example the lower bound $\ln d$ in Ineq.~\eqref{eq:H1MHselfdualbound} is not saturated on KD positive pure states, proving the claim that followed Proposition~\ref{prop:H1MHselfdualbound}.

To further illustrate the complex behaviour of the entropies that can occur with non-MUB KD distributions, we first analyse some further properties of the KD representation associated with the specific $U$ in~Eq.~\eqref{eq:Upet}. We have established numerically that Ineq.~\eqref{eq:graal} holds in this case. For that purpose, we set the $\hat{A}$ basis to be the computational basis and the $\hat B$ basis to be the columns of ${U}$ in Eq. \eqref{eq:Upet}. Then, the KD-positive states must have real density matrices \cite{Langrenez2024Characterizing}. By concavity, we know that the minimum of the KD-Shannon entropy over all KD-positive states is reached on the extremal KD-positive states. To search over the extremal states, we start with the maximally mixed state $\mathbb{I}/3$ and sample a random real direction. We then propagate the state out in this direction until it stops being a state or KD positive.
 We then conduct a guided search around the extremal  positions so detected.    Testing in this manner  $500$ directions with $10,000$ local searches strongly indicates that the minimum is indeed reached on the pure KD-positive states.  Hence, since 
\begin{equation}\label{eq:nice}
    \min_{\mathcal E_+^{\mathrm{KD,pure}}} \left(H_1(\hat A)+H_1(\hat B)\right)=\min_{\mathcal E_+^{\mathrm{KD,pure}}}H_1^{\mathrm{KD}},
\end{equation}
one then easily establishes using subadditivity  and Ineq.~\eqref{eq:graal} that 
\begin{equation}\label{eq:nice2}
\min_{\mathcal E_+^{\mathrm{KD}}}\left(H_1(\hat A)+H_1(\hat B)\right)=\min_{\mathcal E_+^{\mathrm{KD},\mathrm{pure}}}H_1^{\mathrm{KD}}.
\end{equation}
However, for this example, 
\begin{equation}
\label{eq:2lowH1AH1B}
\min_{|\psi\rangle\langle \psi|}\left(H_1(\hat A)+H_1(\hat B)\right)< 
\min_{\mathcal E_+^{\mathrm{KD}}}\left(H_1(\hat A)+H_1(\hat B)\right),
\end{equation}
which we show by establishing numerically that
\begin{equation}\label{eq:lowH1AH1B}
\min_{|\psi\rangle\langle \psi|}\left(H_1(\hat A)+H_1(\hat B)\right) \approx 0.933 < 
\min_{\mathcal E_+^{\mathrm{KD,pure}}} H_1^{\mathrm{KD}} \approx 0.965.
\end{equation}
This implies that the minimal KD-Shannon entropy is strictly larger than the minimal Shannon entropy of $H_1(\hat A)+H_1(\hat B)$, which is thus reached on a KD-nonpositive state. We therefore conclude that the two notions of minimal uncertainty attached to these two different notions of entropy do not coincide in this example. We recall that this phenomenon does not occur for MUB KD representations.

If  we know that the min of $H_1^{\mathrm{KD}}$ on the set of KD positive pure states is reached on basis states, then Eq.~\eqref{eq:nice} is satisfied because, on the basis states, the two entropies are always equal, for all KD representations.  This is precisely what happens in the above example.   So in cases when the only  KD-positive pure states are the basis states, Eq.~\eqref{eq:nice} holds. 
This phenomenon happens with probability one as, for a probability-one subset of transition matrices $U$, the set of KD-positive states is ~\cite{langrenez2024}
\begin{equation}
\mathcal E_+^{\mathrm{KD}}=\mathrm{conv} (\mathcal E_+^{\mathrm{KD, pure}})=\mathrm{conv} \{|a\rangle\langle a|, |b\rangle\langle b|\}.
\end{equation}
  It then follows immediately that, for such $U$, and for all $\hat\rho\in \mathcal E_+^{\mathrm{KD,pure}}$
\begin{equation}
    H_1(\hat A)+H_1(\hat B)=  H_1^{\mathrm{KD}}(\hat\rho),
\end{equation}
so that Eq.~\eqref{eq:nice2} holds.
 In addition, we have numerically confirmed, for $1,000$ Haar random ${U}$s with $d=2,\ldots, 7$, that 
\begin{equation}
\begin{split}
 \min_{\mathcal E_+^{\mathrm{KD}} } H_1^{\mathrm{KD}}= \min_{\mathcal E_+^{\mathrm{KD}}}\left(H_1(\hat A)+H_1(\hat B)\right) >\\ \min_{|\psi\rangle\langle\psi|}\left(H_1(\hat A)+H_1(\hat B)\right),
\end{split}
\end{equation}
where $|\psi\rangle\langle\psi|$ runs through all pure states. We conclude that our numerical data suggest that the property Eq.~\eqref{eq:2lowH1AH1B}, observed for the special example above, in fact holds with probability one for non-MUB KD representations.

For specific non-MUB KD distributions, Ineq.~\ref{eq:graal} can be violated. 
Below, we illustrate this with an example in $d=4$. We define a KD distribution with respect to the transition unitary given by
\begin{align}
U_0 &=\frac{1}{19}\begin{pmatrix}-8&15&-6&-6\\3&-8&-12&-12\\-12&-6&10&-9\\-12&-6&-9&10\end{pmatrix}.
\end{align}
For the KD distribution defined by $U_0$ the set of pure KD-positive states is larger than the set of pure basis states. That is, there are KD-positive pure states that are neither in $\{\ket{a_i}\}$ nor $\{\ket{b_j} \}$. These states exist because the bases defined by $ U_0$ are not completely incompatible \cite{DeBievre2021,DeBievre2023}: the minor of $ U_0$ on rows $\{1,2\}$ and columns $\{3,4\}$ vanishes. Hence $\mathrm{span}\{\ket{a_1},\ket{a_2}\}\cap\mathrm{span}\{\ket{b_1},\ket{b_2}\}\neq\{0\}$. Explicitly,
 \begin{equation} \ket{\psi_{\min}}=\tfrac{1}{\sqrt5}\big(2\ket{a_1}-\ket{a_2}\big)=\tfrac{1}{\sqrt5}\big(-\ket{b_1}+2\ket{b_2}\big), 
\end{equation}
 so that 
 \begin{equation} 
 Q^{\mathrm{KD}}(\psi_{\min})=\frac{1}{95}
\begin{pmatrix}
16&60&0&0\\3&16&0&0\\0&0&0&0\\0&0&0&0
\end{pmatrix}
\geq0 .
 \end{equation} 
Thus $\ket{\psi_{\min}}$ is a KD-positive pure state that is not a basis state. Moreover, $\ket{\psi_{\min}}$ has lower entropy than all the basis states: $H^{\mathrm{KD}}_1(\ket{\psi_{\min}}\bra{\psi_{\min}}) \approx 0.99 < \min_{\hat{\rho} \in \mathcal{E}^{\mathrm{KD,basis}}_+} H^{\mathrm{KD}}_1(\hat{\rho})\approx 1.061$. 

Next, we study the KD distribution defined with respect to a perturbed transition matrix  $  U=  U_0  W$. Here, 
\begin{equation}
  W=\frac{1}{\sqrt{626}}\begin{pmatrix}25&0&-1&0\\0&\sqrt{626}&0&0\\1&0&25&0\\0&0&0&\sqrt{626}\end{pmatrix}
\end{equation}
rotates $\ket{b_1},\ket{b_3}$ by the small angle $\arctan(1/25)\approx0.04$. The top-right minor in $U$ no longer vanishes and no pure state in the vicinity of $\ket{\psi_{\min}}$  is KD positive. There still exist non-basis KD-positive pure states. The minimum KD-Shannon entropy over these pure KD-positive states is reached on a basis state: $H^{\mathrm{KD}}_1(\ket{a_1}\bra{a_1})\approx1.054$. However, the state
\begin{equation}
\hat\rho_*=\frac{1}{374}\begin{pmatrix}303&-142&0&0\\-142&71&0&0\\0&0&0&0\\0&0&0&0\end{pmatrix} ,
\end{equation}
which lies in the vicinity of $\ket{\psi_{\min}}$,
is a rank-two KD-positive state of $ U$. The entropy of the state is $H^{\mathrm{KD}}_1(\hat\rho_*)\approx1.033$.  Hence
\begin{equation}
\min_{\hat\rho\in\mathcal E^{\mathrm{KD}}_+}H^\mathrm{KD}_1(\hat\rho)\leq H^\mathrm{KD}_1(\hat\rho_*)<\min_{\hat\rho\in\mathcal E^{\mathrm{KD},\mathrm{pure}}_+}H^{\mathrm{KD}}_1(\hat\rho),
\end{equation}
violating Ineq.~\ref{eq:graal}.

\section{The case of Wigner-Shannon: continuous variables}\label{app:WignerShannon}
The Wigner function of a state $\hat \rho$ is defined as
\begin{equation}
    W[\hat\rho](q,p)=\frac1\pi \int \exp(2iyp)\langle q-y|\hat\rho|q+y\rangle \rd y.
\end{equation}
In what follows, we will drop the $\hat\rho$ dependence of $W$ in the notation. Then
\begin{equation}
\int_{\R^2}  W(q,p)\rd q\rd p=1
\end{equation}
and
\begin{equation}
\Tr(\hat\rho^2)=2\pi\int_{\R^2} W^2(q,p)  \rd q\rd p.
\end{equation}
The Wigner-Shannon entropy of a Wigner positive state $\hat \rho$ is:
\begin{equation}
H^{ W}_1(\hat\rho)=-\int_{\R^2} W(q,p) \ln  W(q,p) \rd q\rd p.
\end{equation}
By convexity and the Jensen inequality we find that
\begin{equation}\label{WSEnt_basicbound2}
H^{W}_1(\hat\rho)> -\ln\left(\int W^2 \rd q \rd p\right)=\ln(2\pi) -\ln\left(\Tr(\hat\rho^2)\right).
\end{equation}
This inequality, of which a different proof was given in~\cite{Dias2023}, is necessarily strict, since Wigner distributions cannot be constant on their support~\cite{janssen98}. Since the Wigner distribution is Born compatible for $\hat x$ and for $\hat p$, subadditivity of the differential Shannon entropy implies that
\begin{equation}
    H_1(\hat x)+H_1(\hat p)\geq H^{\tilde W}(\hat\rho)>\ln(2\pi)-
    \ln(\Tr(\hat\rho^2)).
\end{equation}
This is the analog of Theorem~\ref{thm:main1}~(i)(Ineq.~\eqref{eq:ineqstringKDGW}), because the Wigner-Weyl representation of quantum mechanics is strongly selfdual with, $\kappa=\kappa_\infty=(2\pi)^{-1}$. However, parts (ii), (iii) and (iv) of Theorem~\ref{thm:main1} do not carry over. 
Indeed, we know from Hudson's theorem~\cite{Hudson1974} that the set of pure Wigner-positive states $\mathcal E_+^{\mathrm W, \mathrm{pure}}$ is given by the pure Gaussian states. Their Wigner-Shannon entropy is readily computed. Indeed,
for the vacuum state, the Wigner function is
\begin{equation}
 W(q,p)=\frac{1}{\pi} \exp(-(q^2+p^2)).
\end{equation}
Hence
\begin{equation}
    H_1^{W}(\hat\rho)=\ln\pi +1>\ln(2\pi). 
\end{equation}
Since $H_1$ is symplectically invariant, the same value is obtained on all pure Gaussian states, and in particular on coherent states. We therefore find that, for all $\hat\rho\in \mathcal E_+^{\mathrm{W},\mathrm{pure}}$,
\begin{equation}
     H_1^{\mathrm W}(\hat \rho)=\ln\pi +1>\ln(2\pi).
\end{equation}
Hence part~(ii) of Theorem~\ref{thm:main1} does not carry over to the present setting.  
For coherent states, since the Wigner function is a product, and because the Wigner distribution is Born compatible for $\hat x$ and for $\hat p$, one finds
\begin{equation}
    H_1(W)=H_1(\hat x)+H_1(\hat p)=\ln\pi+1>\ln(2\pi).
\end{equation}
Since the BBM bound~\eqref{eq:BBM} is saturated only on coherent states~\cite{Bialynicki-Birula1975-ks}, part (iv) of Theorem~\ref{thm:main1} does not hold in the present context. A fortiori, neither does (iii). 

Whereas Theorem~\ref{thm:main1} implies Ineq.~\eqref{eq:graal}, the latter is a weaker statement.  As mentioned in the introduction, it was conjectured that Ineq.~\eqref{eq:graal} holds for the Wigner distribution, namely that, for all $\hat\rho\in \mathcal E^{\mathrm{W}}_+$
\begin{equation}
    H_1^W(\hat\rho)\geq \min_{\hat\rho_+^{\mathrm{pure}}\in \mathcal E_+^{\mathrm{W}, \mathrm{pure}}} H_1^W(\hat\rho_+^{\mathrm{pure}})=\ln(\pi)+1.
\end{equation}
This has been proven to be false recently~\cite{he2026wigner, vanherstraeten2026,  liu2026}.

\bibliography{Bibliography}
\end{document}